%% file: root.tex
\documentclass[a4paper, 10pt, twocolumn]{article}

\usepackage[utf8]{inputenc}         

\usepackage{graphicx}               
\usepackage{tabularx}               
\usepackage{multirow}               
\usepackage{url}                

\usepackage{subcaption}
\usepackage{parskip}
\usepackage{amsmath}                

\usepackage{booktabs}
\usepackage{tabularx}
\usepackage{qrcode}

\newcolumntype{Y}{>{\centering\arraybackslash}X}    

\input{_input/00_preamble.tex}

\begin{document}

\date{}                                         

\title{\vspace{-8mm}\textbf{\large
Reproducible Research is more than Publishing Research Artefacts:\\\vspace{-0.25em}
A Systematic Analysis of Jupyter Notebooks from Research Articles
}}

\author{
    Max Schröder$^{1,2}$, Frank Krüger$^1$, and Sascha Spors$^1$\\
    $^1$ \emph{\small Institute of Communications Engineering, University of Rostock}\\
    $^2$ \emph{\small University Library, University of Rostock}\\
    \emph{\small E-Mail: \{max.schroeder, frank.krueger, sascha.spors\}@uni-rostock.de}
} \maketitle
\thispagestyle{empty}           

\section{Introduction}
\label{sec:intro}\vspace{-10pt}


With the advent of Open Science, researchers have started to publish their \emph{research artefacts} (\ie data, software, and other products of the investigations) in order to allow others to reproduce their investigations.
While this publication is beneficial for science in general \cite{McKiernan2016}, it often lacks a comprehensive documentation and completeness with respect to the artefacts.
This, in turn, prevents the successful reproduction of the analyses.

\begin{figure}[t!]
    \centering
    \includegraphics[width=.48\textwidth,frame]{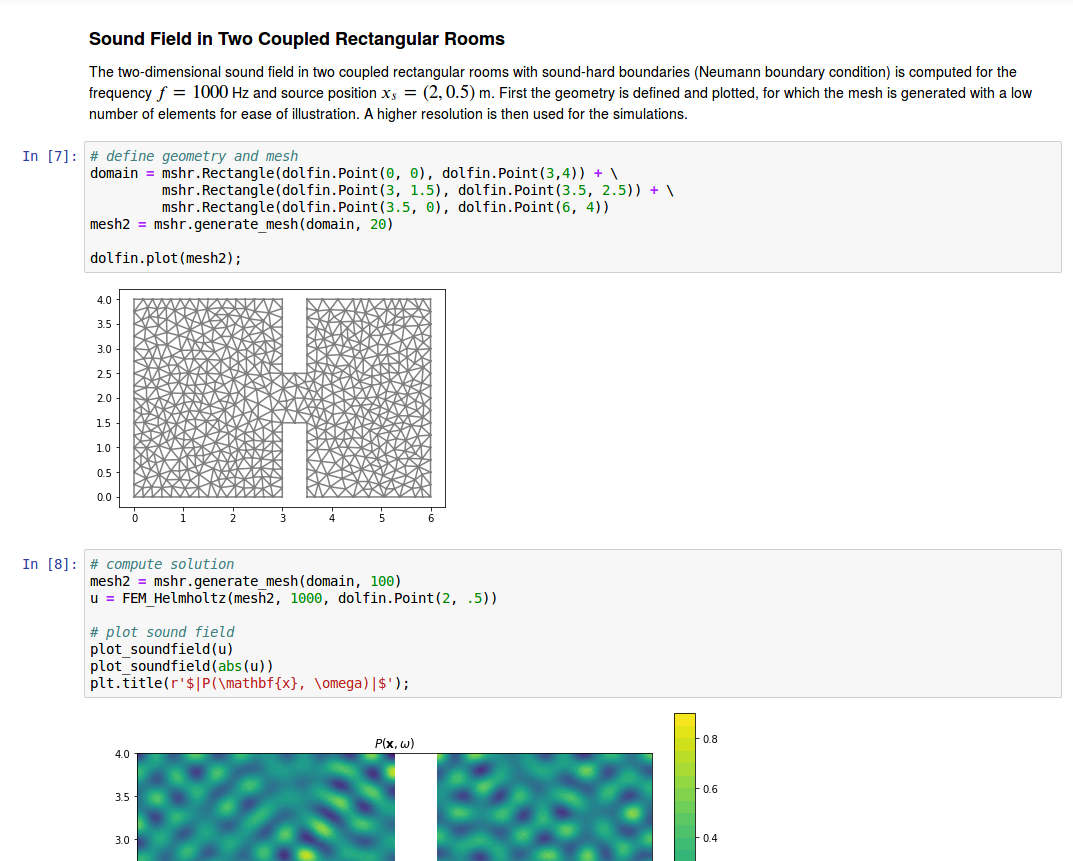}
    \caption{Screenshot of a Jupyter notebook \cite{Spors2018} containing three `cells': the first one contains textual documentation and the latter two contain python source code followed by the corresponding output in the form of plots.}
    \label{fig:jupyter}
\end{figure}

\emph{Jupyter notebooks} recently have gained increased attention as a method for publishing research investigations \cite{Perkel2018, Shen2014}.
The term `Jupyter notebook' is used for both:
\begin{inparaenum}[(1)]
    \item a JSON document (*.ipynb files) and
    \item a corresponding web service for reading, writing, and executing these notebooks.
\end{inparaenum}
The notebooks encapsulate both documentation and source code along with the source code output \ie the results in so-called \emph{cells} inside a single streamlined document (see \figref{jupyter}).
The web application enables text editing, but also the interactive execution of the source code cells and, thus, the re-execution of the previous investigations.
For this purpose, different programming kernels such as Python, R, and Octave exists.

Despite the increasing use of Jupyter notebooks for the publication of research investigations, their reproducibility is not automatically guaranteed by the document format and the web service.
A crucial reproducibility requirement is the detailed documentation of the \emph{computing environment}.
As new software releases are being developed, the dependencies have to explicitly state the software version used for the study.
Ideally, the environment is ready-to-use for other researchers by containerisation techniques.

We have been \emph{systematically} analysing research publications that also published their investigations as Jupyter notebooks.
In this paper, we present preliminary results of this analysis for five publications.
The results show, that the quality of the published research artefacts must be improved in order to assure reproducibility.

The next section introduces how relevant publications were identified and how the analysis was performed.
The results are described in \Secref{results} which is followed by our conclusion (\secref{conclusion}).

\vspace{-10pt}
\section{Method}
\label{sec:method}\vspace{-10pt}

\begin{figure*}[t]
    \centering
    \begin{subfigure}[t]{0.5\textwidth}
        \centering
        \includegraphics[width=.6\textwidth]{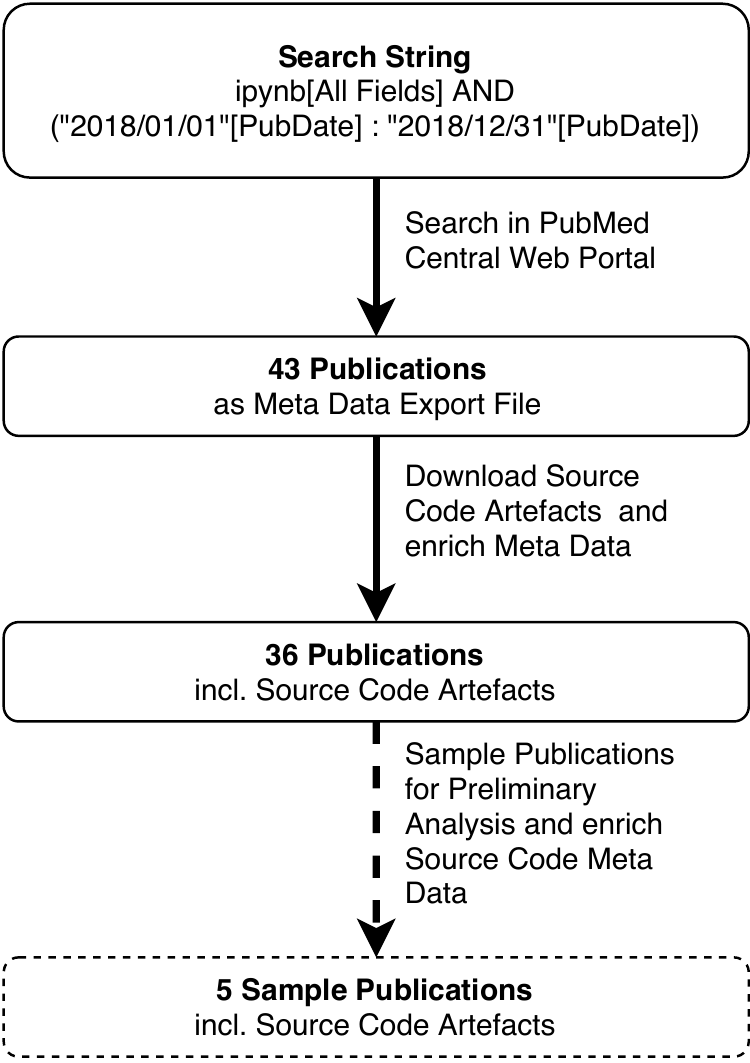}
    \end{subfigure}%
    ~
    \begin{subfigure}[t]{0.5\textwidth}
        \centering
        \includegraphics[width=.7\textwidth]{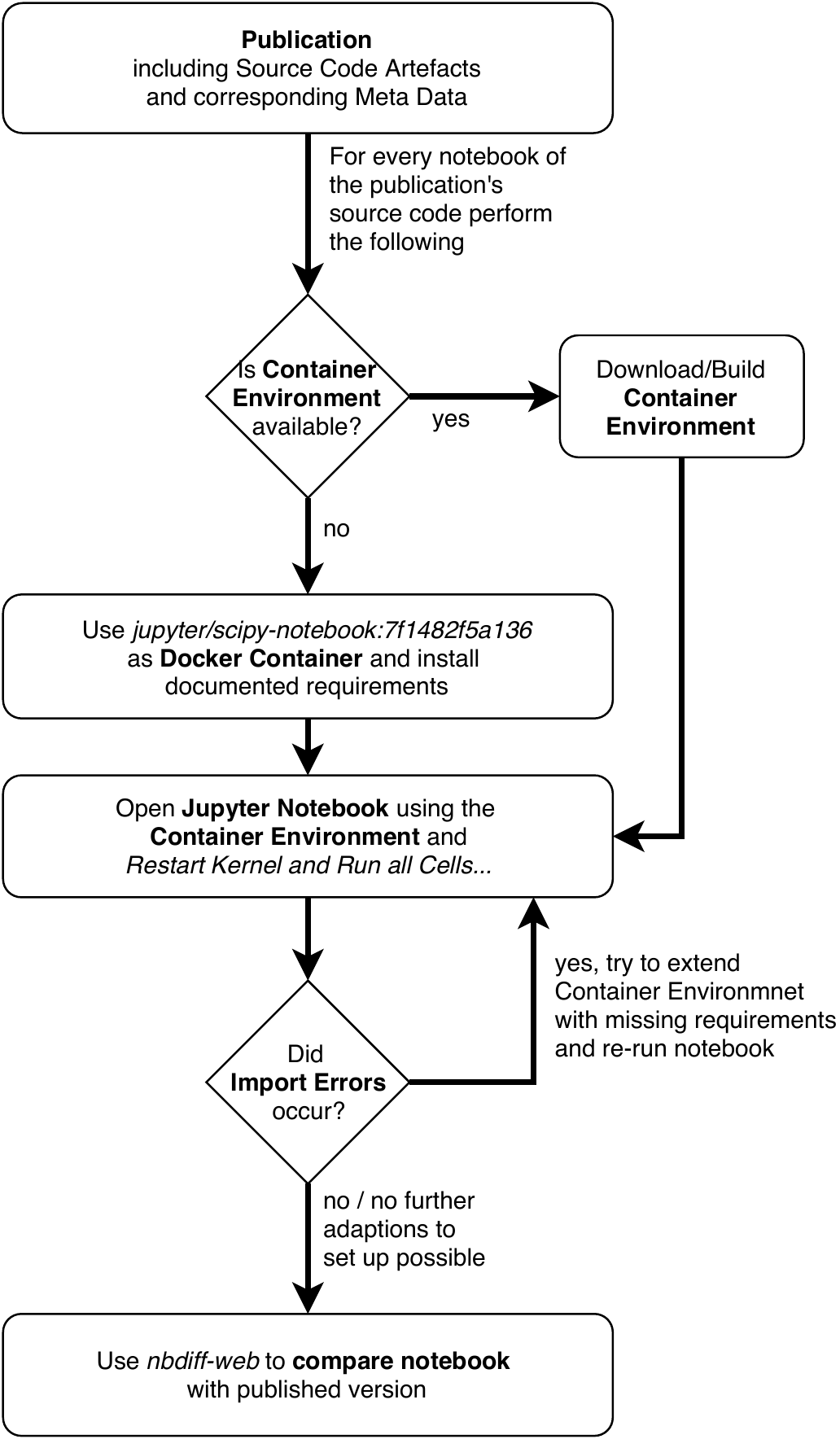}
    \end{subfigure}
    \caption{The identification workflow of relevant publications from the PubMed Central database (left); the reproducibility analysis performed for every identified publication (right).}
    \label{fig:workflows}
\end{figure*}

In order to identify recent publications including their published Jupyter notebooks, we used the PubMed Central database (PMC) \cite{Roberts381}.
This database has been chosen, because its articles are freely accessible and it contains a large number of journals.
The notebook file type `ipynb' was used as search term.
Additionally, we restricted the results to publications from the year 2018 as we expect that software dependencies are more likely to be available and that better practices are implemented in more recent publications.
From 43 publications in the PMC database seven publications were excluded as the jupyter notebooks mentioned do not refer to the original analysis of the publication (n=6) or are no longer available in the repository (n=1).
For the preliminary results in this paper, we randomly selected five publications from the resulting 36 publications.
\Figref{workflows} (left) illustrates the overall workflow.

Each of the five publications was analysed with respect to the reproducibility of its jupyter notebooks as specified in \Figref{workflows} (right).
If the computing environment is not available as virtual machine or container image, we attempted to reconstruct it based on the documentation within the repository.
For this purpose, we use the containerisation solution `Docker' and an official Jupyter notebook image that is incrementally extended by installing requirements.
The reconstruction attempt was limited to 3 hours and a single environment per article, as we with unlimited resources the reconstruction is always possible and authors use a homogeneous environment during their investigations if not stated otherwise.
Afterwards, all cells of the Jupyter notebook are executed and the result compared to the published version using the notebook differentiation tool `nbdime'.
The analysis in this paper is performed on a Dell Latitude 7490 with Ubuntu\footnote{see \url{https://www.ubuntu.com/}} v16.04.6, Docker\footnote{see \url{https://www.docker.com/}} v18.09.3, and conda\footnote{see \url{https://anaconda.org/}} v4.6.8.

\vspace{-10pt}
\section{Results}
\label{sec:results}\vspace{-10pt}

The meta data of the overall set of relevant publications is analysed in the next section.
\Secref{results_jupyter} contains the preliminary results of a detailed evaluation of the reproducibility of the Jupyter notebooks.

\vspace{-10pt}
\subsection{Meta Data Analysis}
\label{sec:results_meta}\vspace{-10pt}

\begin{figure}[t]
    \centering
    \includegraphics[width=.4\textwidth]{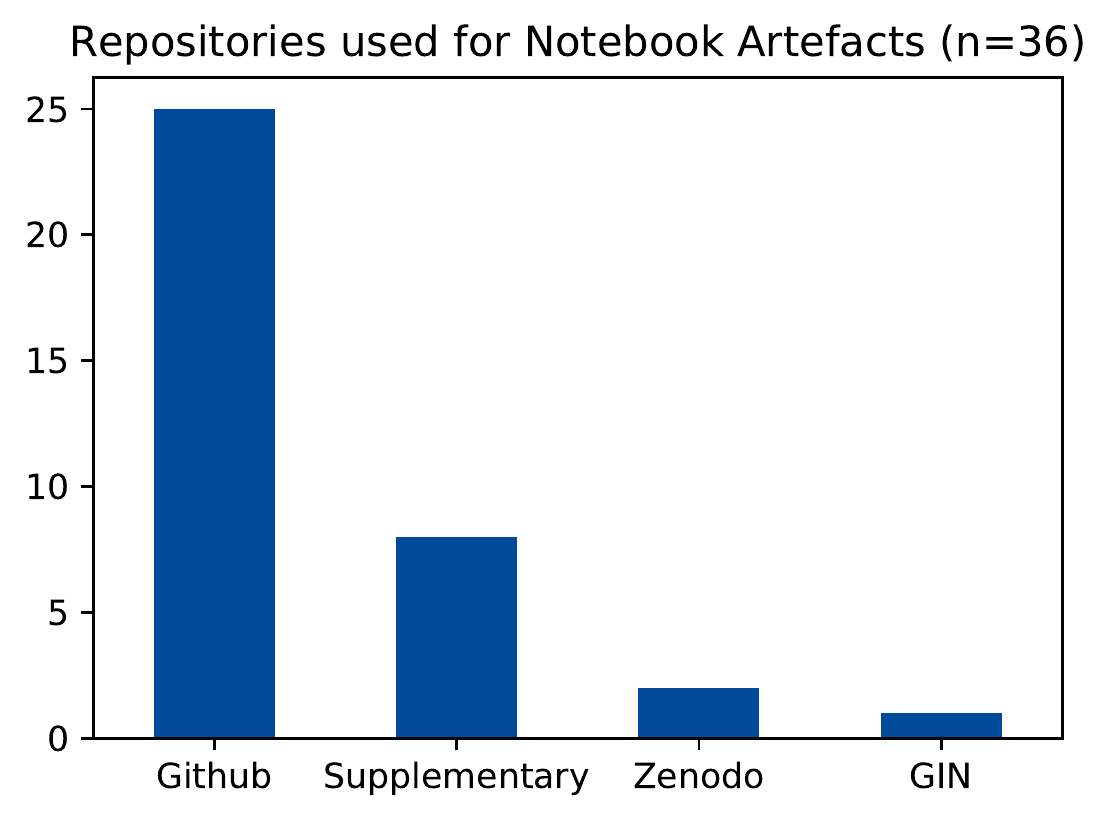}
    \caption{Which repositories are used to publish source code artefacts \ie Jupyter notebooks?}
    \label{fig:result_repositories}
\end{figure}

The meta data has been analysed for the full set of identified publications (n=36) according to \figref{workflows} (left).
We evaluated two aspects that are crucial for the accessibility of the source code artefacts: the repository that is used and the source code license (consecutively):

GitHub is the most frequently used repository for the upload of source code artefacts (see \figref{result_repositories}).
Most mentions of GitHub repositories in the publications, however, lack in specifying a concrete version of the source code that is used for the investigations and mention the base url only.
Although this is common practice especially for software projects that are continuously developed, this prevents others from comprehending the original study.
Instead, the other repositories Supplementary Material, Zenodo\footnote{see \url{https://zenodo.org/}}, and GIN\footnote{see \url{https://web.gin.g-node.org/}} reference the original artefacts.

\begin{figure}[t]
    \centering
    \includegraphics[width=.48\textwidth]{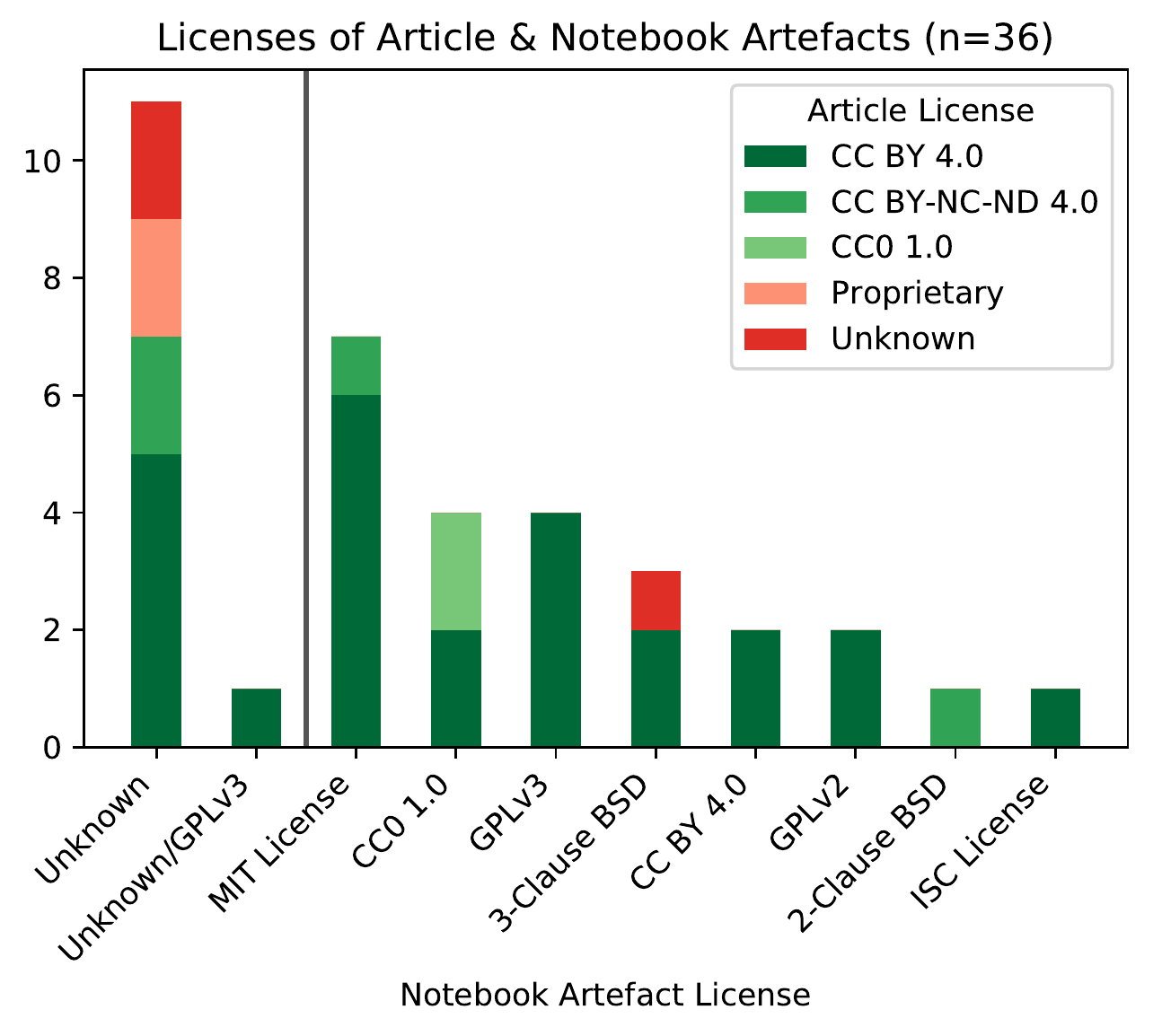}
    \caption{
        Shows the publishing license of the Jupyter notebooks \wrt the publishing license of the article.
        The vertical line separates problematic artefact licenses (left) from standardized open access licenses (right).}
    \label{fig:result_licenses}
\end{figure}

Even though the publication is publicly accessible by a proper license, this needs not be true for the source code artefacts that have to be declared as well.
Unfortunately, almost one third of the source code artefacts lack in providing a license for their usage (see \figref{result_licenses}) and, thus, impeding other researchers from re-using their investigations.
For the software artefacts the most common license is the MIT\footnote{see \url{https://opensource.org/licenses/MIT}} license, the GPLv3\footnote{see \url{https://www.gnu.org/licenses/gpl-3.0}}, and interestingly the CC0 1.0\footnote{see \url{https://creativecommons.org/publicdomain/zero/1.0/}} which gives all usage rights to the public domain.
Interestingly, some of the articles even though listed in the PMC database are not freely accessible, but only via proprietary licenses or with no license information at all.
One software artefact had no license information within the repository branch mentioned in the publication, but within the master branch the license was declared as GPLv3.
In \Figref{result_licenses} this is depicted as `Unknown/GPLv3'.

\input{_input/notebooks}

\vspace{-10pt}
\subsection{Reproducibility Analysis}
\label{sec:results_jupyter}\vspace{-10pt}

For the preliminary reproducibility analysis, we randomly selected five publications from the overall set of publications (see \secref{method}).
Each of these publications has been evaluated in detail \wrt the following reproducibility aspects:

\newcommand\rref[1]{R\ref{r:#1}}%
\newcommand\rlabel[1]{\label{r:#1}}%

\begin{compactenum}[{R}1.]
    \item \rlabel{mention}
        How many notebooks are mentioned and how many are published?
    \item \rlabel{documentation}
        Where are the source code artefacts documented beside the publication?
    \item \rlabel{requirements}
        Where can the documentation of the software requirements be found?
    \item \rlabel{environment}
        Is the computing environment available or can it be reconstructed from the documentation?
    \item \rlabel{data}
        Is the complete raw data of the study available?
    \item \rlabel{reproducibility}
        Can the Jupyter notebook be completely re-executed with the same results?
\end{compactenum}

\begin{figure}[t!]
    \centering
    \includegraphics[width=.48\textwidth]{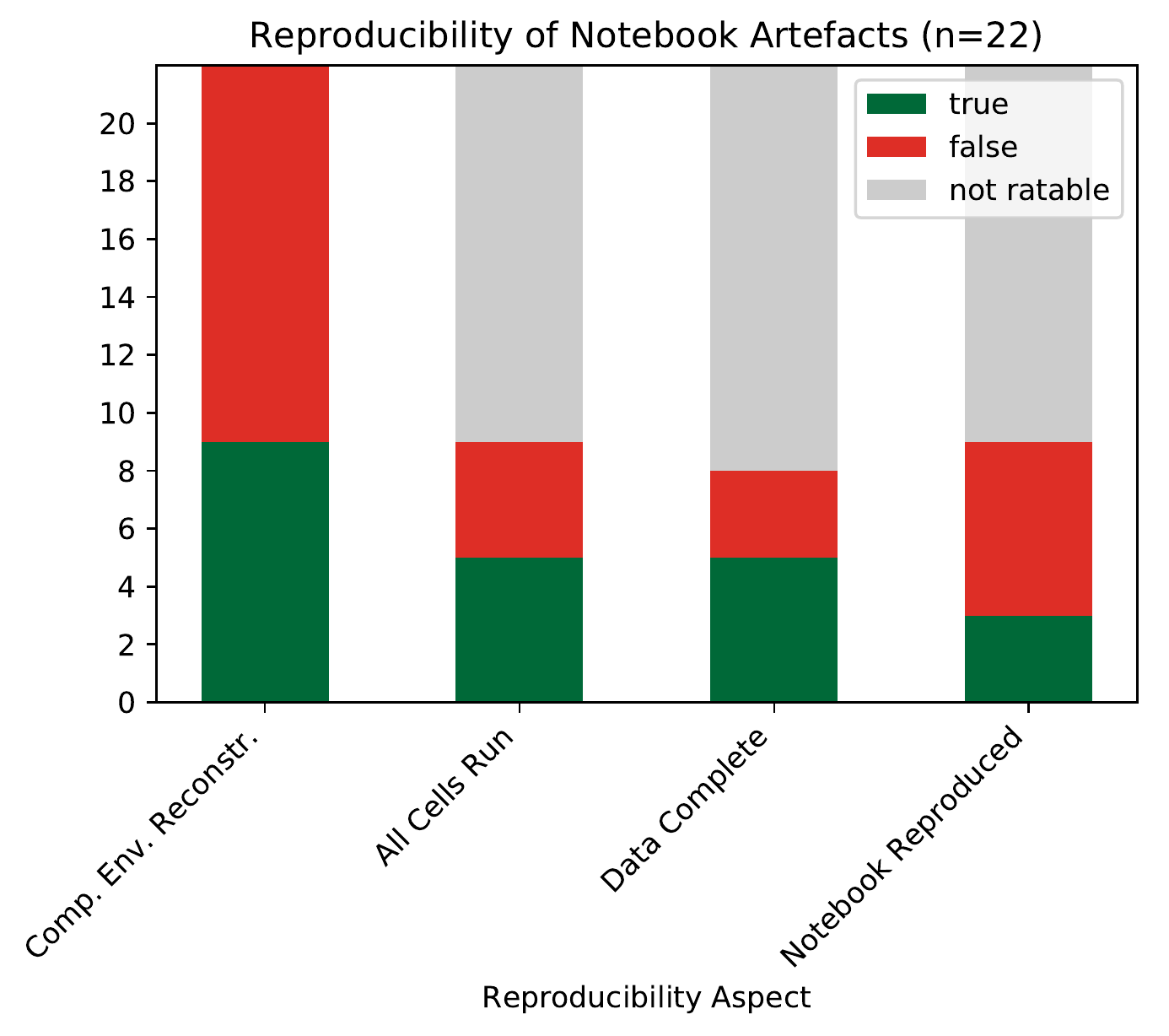}
    \caption{
        Illustration of the reproducibility analysis showing several aspects that were analysed for every notebook from the five sample publications.
        Some aspects could not be rated as prerequisites are not satisfied e.g., a computing environment exists.
    }
    \label{fig:result_reproducibility}
\end{figure}

Interestingly, one repository contained eleven Jupyter notebooks whereas the corresponding article mentioned only a single one (see \tblref{notebooks}).
The other articles mentioned the majority of their notebooks which might indicate, that the publication of the source code is specifically tailored for that article (\rref{mention}).

The documentation of the Jupyter notebooks was not evaluated \wrt their quality but only the presence of comments and explanations (\rref{documentation}).
\Tblref{notebooks} shows that indeed the notebooks included some part of the documentation as well as a readme file.
However, only two projects used methods such as Read the Docs respectively Sphinx\footnote{see \url{https://readthedocs.org/}} and mkdocs\footnote{see \url{https://www.mkdocs.org/ }} that are specifically tailored for extended documentation.
Interestingly, the documentation of the requirements (\rref{requirements}) can not always be found in the notebook itself.
Other sources of documentation are also containerisation descriptions (see \tblref{notebooks}).

Another essential aspect for re-using Jupyter notebooks is that a computing environment is available that behaves identical \wrt the application compared to the environment used by the authors (\rref{environment}).
Environment here denotes the software dependencies that are needed to execute the notebook.
Unfortunately, only one project used containerisation techniques such as Docker and Travis-CI\footnote{see \url{https://travis-ci.org/}} so that for all other projects the environment has to be reconstructed from scratch (see \tblref{notebooks}).
We call a reconstruction of the environment successful if no import errors arose.
This reconstruction was successful for 9 from the 22 notebooks from the five sample publications (see \figref{result_reproducibility}).
The majority of environments where the reconstruction was unsuccessful results from the software artefact with 11 notebooks even though this repository contained descriptions of container images.
The reason was that the container could not be build and no public accessible image was found.
Although the reconstruction for the other notebooks was successful in most cases, the documentation often lacks in providing a complete list of python packages including their required version requiring more effort during the reconstruction (\Tblref{notebooks}).

For those Jupyter notebooks where a computing environment could be reconstructed, we re-executed the notebook by using the build-in mechanism `Restart Kernel and Run all Cells...'.
All cells run successfully for only five notebooks (see \Figref{result_reproducibility}).
However, three notebooks failed as the data was not available (\rref{data}).
The remaining notebooks failed due to an unset constant.

From the 22 notebooks only three notebooks could be successfully reproduced (\rref{reproducibility}) meaning that not only all cells run successfully but also the output is equal to the originally published version.
One of the three notebooks, however, contained function definitions only.

\vspace{-10pt}
\section{Conclusion}
\label{sec:conclusion}\vspace{-10pt}

In this paper we showed preliminary results of a systematic reproducibility analysis that indicate potential improvements when publishing research investigations in the form of Jupyter notebooks.
Our findings are substantiated by the analysis of Chen \etal \cite{Chen2018} who analyse reproducible research in the field of high-energy physics.
While we strongly support the idea of Open Science, we think that research artefact publications must be cared as much as for the article itself in order to enable re-usage.
Rule \etal \cite{Rule2018} provide ten simple rules for reproducible research with Jupyter notebooks that is consistent with the problems we identified in this analysis and, thus, can help researchers in providing reproducible Jupyter notebooks.

\vspace{-10pt}
\section*{Acknowledgement}
\label{sec:acknowledgement}\vspace{-10pt}

This research was supported by the German Research Foundation (Deutsche Forschungsgemeinschaft, DFG) within the Collaborative Research Centre 1270 ELAINE.

{
    \footnotesize
    \bibliographystyle{IEEEtran}
    \bibliography{_references/jupyter}
}
\end{document}

%% file: _input/00_preamble.tex
\usepackage{booktabs}
\usepackage{tabularx}
\usepackage{caption}
\usepackage{listings}
\usepackage{xcolor}
\usepackage{xspace}
\usepackage{paralist}
\usepackage{amsmath,amsfonts,amssymb,stmaryrd}
\usepackage{bigints}
\usepackage{graphicx}
\usepackage{oz}
\usepackage{ifthen}
\usepackage{tabularx}
\usepackage{helvet}
\usepackage{algpseudocode}
\usepackage{algorithm}
\usepackage{algorithmicx}
\usepackage{multirow}
\usepackage{subcaption}
\usepackage[export]{adjustbox}
\usepackage{csquotes}

\usepackage{tikz}
\usepackage{pgfplots}
\usetikzlibrary{external}
\usepackage{textcomp}
\usetikzlibrary{shapes,arrows, positioning}

\usepackage{hyperref}

\algrenewcommand\alglinenumber[1]{\tiny #1:}

\newtheorem{cexpl}{Example}

\let\oldcexpl\cexpl
\renewcommand{\cexpl}{\oldcexpl\normalfont}

\newcommand{\etal}{et~al.\xspace}
\newcommand{\wrt}{w.r.t.\ }
\newcommand{\ie}{i.\,e.,\xspace}

\newcommand{\secref}[1]{Section~\ref{sec:#1}}
\newcommand{\Secref}[1]{Section~\ref{sec:#1}}
\newcommand{\figref}[1]{Fig.~\ref{fig:#1}}
\newcommand{\Figref}[1]{Figure~\ref{fig:#1}}
\newcommand{\tblref}[1]{Table~\ref{tbl:#1}}
\newcommand{\Tblref}[1]{Table~\ref{tbl:#1}}

\definecolor{actcolor}{RGB}{159,177,252}
\definecolor{entcolor}{RGB}{255,252,135}
\definecolor{usecolor}{RGB}{139,0,0}
\definecolor{gencolor}{RGB}{0,100,0}
\tikzset{%
		activity/.style    = {draw, thick, rectangle, minimum height = 2em,fill=actcolor},
		entity/.style      = {draw, circle, minimum height = 2em, fill=entcolor}, 
	}

%% file: _input/notebooks.tex
\setlength{\tabcolsep}{12pt}

\begin{table*}[t!]
    \centering
    \begin{tabular}{ccllll}
    \toprule
    \textbf{Ref.} & \textbf{Mentions} &                        \textbf{Documentation} &                           \textbf{Requirements} &    \textbf{Req. Problems} \\
    \midrule
    \cite{Tambe2018}          &    1 / 2 &  notebook, readme     & no documentation     & --- \\
    \cite{Cummins2018}        &   1 / 11 &  html, notebook,      & docker, notebook, readme & docker fails build, custom\\
                            &          &  readme, readthedocs  & readthedocs, travisci & image did not install\\
                            &          &                       &              &  successfully\\
    \cite{Yang2018}           &    1 / 2 &  mkdocs, notebook,    & mkdocs               & missing versions \\
                            &          &  readme\\
    \cite{Collier2018}        &    4 / 6 &  notebook, notes,     & notebook, readme     & missing python \\
                            &          &  readme               &                      & packages\\
    \cite{Fisher-Wellman2018} &    1 / 1 &  notebook, readme     & no documentation     & --- \\
    \bottomrule
    \end{tabular}
    \caption{
        Summary of the Notebook Meta Data analysis for the five sample publications:
        `Mentions' refers to the number of jupyter notebook mentions within the publication compared to the number of jupyter notebooks within the source code repository; `Req. Problems' refers to problems with documented requirements.
    }
    \label{tbl:notebooks}
\end{table*}